\documentclass[aps,prl,twocolumn,groupedaddress]{revtex4}

\usepackage{epsf}
\usepackage{graphicx}% Include figure files
\usepackage{dcolumn}% Align table columns on decimal point
\usepackage{bm}% bold math

\begin{document}

%Title of paper
\title{Quantum lifetime of 2D electron in magnetic field}

\author{Scott Dietrich}
\author{Sergey Vitkalov}
\email[Corresponding author: ]{vitkalov@sci.ccny.cuny.edu}
%\homepage[]{Your web page}
%\thanks{}
%\altaffiliation{}
\affiliation{Physics Department, City College of the City University of New York, New York 10031, USA}
\author{D. V. Dmitriev}
\author {A. A. Bykov}
\altaffiliation{Novosibirsk State Technical University, 630092 Novosibirsk, Russia}
\affiliation{Institute of Semiconductor Physics, 630090 Novosibirsk, Russia}

\date{\today}

\begin{abstract} 
The lifetime of two dimensional electrons in GaAs quantum wells, placed in weak quantizing magnetic fields, is measured using a simple transport method in broad range of temperatures  from 0.3 K to 20 K. The temperature variations of the electron lifetime are found to be in good agreement with conventional theory of electron-electron scattering in 2D systems.
\end{abstract}
  

\maketitle

\section{Introduction}

Transport properties of 2D electrons, placed in quantizing magnetic fields, attract a great deal of attention for its fundamental significance. Despite a long history \cite{shoenberg1984}, the research in this direction is still active and very surprising. Significant progress has been made during the last decade in understanding the nonlinear properties of the 2D electrons in crossed electric and magnetic fields.   In response to both microwave radiation and $dc$ excitations, strongly nonlinear electron transport \cite{zudov2001,engel2001,yang2002,dorozh2003,willett2004,mani2004,kukushkin2004,stud2005,bykov2005,bykovJETP2006,bykov2007R,zudov2007R,du2007,stud2007,zudovPRB2008,bykov2008,bykov2008a,bykov2008b,gusev2008a,zudov2009,hatke2009a,dorozh2009,vitkalov2009,vitkalov_review2009,gusev2009,durst2003,ryzhii1970,anderson,shi,liu2005,dietel2005,inarreaPRB2005,vavilov2004,dmitriev2005,alicea2005,volkov2007,glazman2007,dmitriev2007} that gives rise to unusual electron states \cite{mani2002,zudov2003,zudov2007,bykov2007zdr,zudov2008zdr,bykov2010,zudov2010,gusev2011,andreev2003,auerbach2005} has been reported and investigated. Very recent experimental studies of the strongly nonlinear resistance in the low frequency domain \cite{bykov2007R,vitkalov2009,gusev2009,vitkalov_review2009} show that the dominant mechanism of the nonlinearity is related to a peculiar quantal heating ("inelastic" mechanism \cite{dmitriev2005}), which may not increase the broadening of electron distribution ("temperature") in systems with discrete spectrum \cite{vitkalov2009,vitkalov_review2009}. Microwave studies of this nonlinearity \cite{hatke2009a} indicate the relevance of another nonlinear mechanism:  electric field induced variations in the kinematics of electron scattering on impurities  ("displacement" mechanism \cite{ryzhii1970,durst2003,vavilov2004}), which limits the lifetime of  an electron in a quantum state.   

The electron quantum lifetime $\tau_q$ has been measured in many experiments \cite{shoenberg1984} and is an important property of two dimensional systems \cite{hamilton2009}. The standard transport method to measure the electron lifetime is based on an extraction of the Dingle factor from temperature or magnetic field dependences of the amplitude of Shubnikov-de-Haas (SdH) oscillations. This method works well at low temperatures, at which the Dingle factor is nearly temperature independent\cite{sdh}. Application of this method to higher temperatures is considered to be problematic, since it involves the separation of unknown, small temperature variations of the Dingle factor from strong variations of SdH amplitude, which is limited by the thermal broadening of the electron distribution. Finally at very high temperature, the SdH oscillations are absent and, thus, the standard method is not applicable.

Recently several transport methods have been introduced to access the temperature dependence of the quantum electron lifetime $\tau_q$. Electric field \cite{zudov2009} and microwave \cite{hatke2009a} induced magnetoresistance oscillations show that the amplitude of the oscillations depends on temperature. At  $T>2K$ variations of the quantum scattering time $\tau_q$ are found to be temperature dependent. At temperature below 2K the dependence saturates, indicating an electric-field-induced overheating. The overheating may create very peculiar electron distributions \cite{vitkalov2009} and is a restriction of this method. In another set of experiments the temperature dependence of the quantum lifetime $\tau_q$ is extracted from the amplitude of magneto-intersubband oscillations (MISO) in double GaAs quantum wells\cite{gusev2008b} and in wide GaAs quantum wells with two occupied subbands \cite{bykov2009}. This method  works in the linear response regime (at thermal equilibrium) but requires two sets of Landau levels and, therefore, does not work for common electron systems with only one populated subband(see also Ref.\cite{berk1995}). The quantum electron lifetime $\tau_q$ was recently accessed   through the amplitude of magnetophonon resistance oscillations \cite{zudov2009ph,bykov2009ph}. However the method depends on the rate of electron-phonon scattering and, therefore, requires  high temperatures.   

This paper presents a simple transport method to measure the quantum electron lifetime $\tau_q$. This method works at thermal equilibrium for common electron systems and is applicable for a broad range of temperatures. The method is based on measurements of a positive magnetoresitance, which is induced by a quantized (periodic) motion of electrons in magnetic fields. Due to the circular motion, a scattered electron may return  to the same impurity repeatedly, enhancing the total scattering amplitude. The stronger the magnetic field, the larger the probability for the electron to return to the same impurity. Thus, the scattering rate increases with the magnetic field, giving rise to the positive magnetoresistance. Quantitative accounts of these quantum interference effects have been done recently\cite{vavilov2004}. 

This paper shows a comparison of the positive magnetoresistance, observed in high mobility GaAs quantum wells, with theory \cite{vavilov2004}. The quantum electron lifetime $\tau_q$ is extracted from the comparison. The time decreases with increasing temperature. In the temperature interval 0.3K -15 K the change of the quantum scattering rate $\Delta (1/\tau_q)=1/\tau_q(T)-1/\tau_q(T=0.3K)$ is proportional to $T^2$. The behavior indicates a dominant contribution of  electron-electron scattering to the decrease of the quantum electron lifetime $\tau_q$ with temperature. The data is in good agreement with theory.

Both positive and negative magnetoresistance  were observed in 2D electron systems \cite{storm,renard,wang,fletcher,kunts},  however the quantum contribution to the magnetoresistance \cite{vavilov2004} has not been identified in those works.  In a very recent experiment on a two-subband electron system, the quantum contribution to the resistance \cite{raichev}  was separated from the classical magnetoresistance \cite{mamani}.

\section{Experimental Setup}

Our samples are high-mobility GaAs quantum wells grown by molecular beam epitaxy on semi-insulating (001) GaAs substrates. The width of the GaAs quantum well is 13 nm. Two AlAs/GaAs type-II superlattices grown on both sides of the well served as barriers, providing a high mobility of 2D electrons inside the well at a high electron density\cite{fried1996}. One sample was studied with electron density $n$ = 8.2 $\times 10^{15}$ m$^{-2}$ and mobility $\mu$= 93 m$^2$/Vs. Another sample with comparable parameters shows similar results. In this paper we show results for the first sample.

The studied 2D electron system is etched in the shape of a Hall bar. The width and the length of the measured part of the sample are $d=$50$\mu m$ and $L=$250$\mu m$. A 12 Hz alternating current is applied through current contacts formed in the 2D electron layer. The longitudinal $ac$ voltage $V_{xx}$ is measured between potential contacts displaced 250$\mu m$ along each side of the sample. The Hall voltage $V_{xy}$ is measured between potential contacts displaced 50$\mu m$ across the electric current.

The current contacts are sufficiently separated from the measured area by a distance of 500$\mu m$, which is much greater than the inelastic relaxation length of the 2D electrons $L_{in}=(D \tau_{in})^{1/2} \sim 1-5 $$\mu m$. This ensures that the experiments are done at thermal equilibrium and the distribution of the electric current is uniform across the samples.  The longitudinal and Hall voltages were measured simultaneously, using two lockin amplifiers with 10 M$\Omega$ input impedances. The potential contacts provided insignificant contribution to the overall response due to small values of the contact resistance (about 1k$\Omega$) and negligibly small electric current flowing through the contacts.

Measurements were carried out for discrete temperature values in the range of 0.3 to 20 Kelvin in a He-3 insert in a superconducting solenoid. Samples and a calibrated thermometer were mounted on a cold copper finger in vacuum. Magnetic fields were applied perpendicular to the 2D electron layers and sweeps were made at each temperature over the range of zero to 0.7 Tesla.

\section{Results and discussion}

\begin{figure}[t!]
\includegraphics[width=3in]{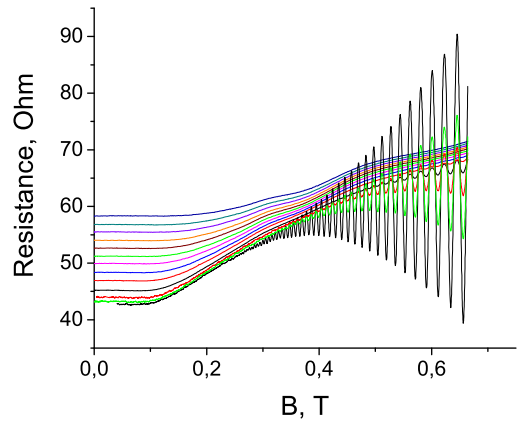}
\caption{(Color online) Resistance versus magnetic field at different temperatures  from the bottom to the top:  2.25K, 2.98K, 3.75, 4.71K, 5.94, 6.95K, 7.96K, 8.85K, 9.80K, 10.73K, 11.75K, 12.66K, and 13.70K.}
\label{exp}
\end{figure}

Figure {\ref{exp}} shows the magnetoresitance of 2D electrons taken at different temperatures. All curves demonstrate a similar behavior. At small magnetic fields ($B<$0.1 T) the curves show extremely weak (unrecognizable in the present scale) dependencies on the magnetic field. At higher magnetic fields, the resistance increases. Not shown in figure 1 is the trace at the lowest temperature T=0.3K, which indicates that the resistance increase correlates with the quantization of the electron spectrum. Namely, the positive magnetoresistance (taken at $T>$2K) starts at the magnetic field at which the quantum (SdH) oscillations are first observed at $T=$0.3 K. The positive magnetoresistance and its temperature dependence are the main targets of the experiments.  Below we compare the resistance increase with the interference enhancement of impurity scattering in the quantizing magnetic fields \cite{vavilov2004}. At even higher magnetic fields the magnetoresistance shows quantum  oscillations, which depend strongly on the temperature. The oscillations are beyond the scope of this paper.

\begin{figure}[t!]
\includegraphics[width=3in]{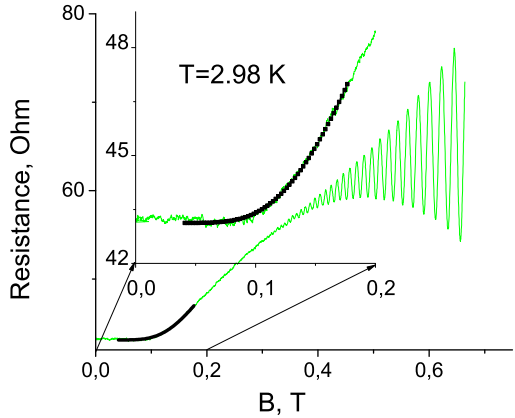}
\caption{(Color online) Data for T=2.98K shown in green (grey) and theoretical curve (eq.(\ref{main})) shown in black.}
\label{comp}
\end{figure}

Figure {\ref{comp}} shows the positive magnetoresistance in better detail and demonstrates a comparison with theory \cite{vavilov2004}. The theory considers 2D electrons, which are moving in magnetic field and scattered by a disordered potential. 
Due to circular electron motion in magnetic field, electrons scattered by an impurity  may return to the impurity again and again, enhancing the total scattering amplitude. The stronger the magnetic field, the more probable it is for the electron to return to the same impurity. A quantitative account of the interference of quantum amplitudes, corresponding to different electrons returns, shows the positive magnetoresistance \cite{vavilov2004}.  Presented in figure \ref{comp}, theoretical dependence is plotted in accordance with the following theoretical expression (see eq. (4.9) in the Ref.\cite{vavilov2004}):

\begin{equation}
R(B)=R_D \cdot [1+2(e^{-\alpha}+ e^{-2\alpha}(1-\alpha)^2)]
\label{main}
\end{equation}

, where $\alpha/2=\gamma/B=\pi/(\omega_c\tau_q)$. The parameter $\gamma$ is a fitting parameter and $R_D$ is a resistance at zero magnetic field \cite{prefactor}. 

Although the theory is developed for a broad range of magnetic fields, in this paper the comparison is done at small magnetic fields, at which the magnetic length $\lambda$ is larger the correlation length of the disorder $\xi$. At this condition the self-consistent Born approximation (SCBA), utilized in the theory for the disordered potential, is accurate and the comparison with the theory is valid. At higher magnetic fields the SCBA approximation fails and corrections to the theory are expected (see below). Furthermore, at high temperatures and magnetic fields $B>$0.3 T the magnetoresistance shows  additional phonon-induced oscillations \cite{zudov_ph,bykov_ph} (see fig. \ref{exp}). These oscillations are also beyond the theory. Finally, the exponentially strong enhancement of the scattering rate at the small magnetic fields provides an immunity of the utilized procedure with respect to possible smooth resistance variations of yet unidentified origin. 

\begin{figure}[t!]
\includegraphics[width=3in]{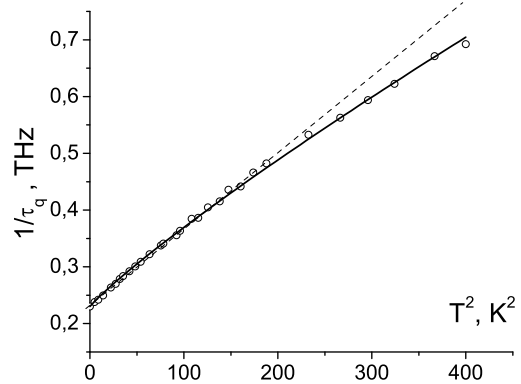}
\caption{Temperature dependence of the quantum scattering rate $1/\tau_q$ (open circles) plotted vs $T^2$. Dashed line presents a linear fit of the temperature dependence at $T<15$K.  Solid line presents expected temperature dependence due to electron-electron scattering (see eq.(\ref{ee})).}
\label{tau_q}
\end{figure}
      
Presented in figure \ref{comp}, comparison with the theory yields the quantum scattering time (electron lifetime) $\tau_q$. Figure \ref{tau_q} shows the temperature dependence of the quantum scattering rate $1/\tau_q$. The dependence is obtained from the comparison of the theory with the magnetoresitance, taken at different temperatures. The dependence is plotted vs the square of the temperature. The plot indicates quadratic variations of the quantum scattering rate $1/\tau_q$ with the temperatures below 15 K. Shown in the figure the dashed straight line approximates the $T^2$ dependence: 

\begin{equation}
1/\tau_{q}(GHz)=234+1.33 \cdot T^2(K^2)
\label{tau_q-T}
\end{equation}

The $T^2$ dependence suggests that the electron-electron scattering is the dominant mechanism, reducing the electron lifetime. In accordance with conventional theory at zero magnetic field \cite{chaplik,quinn,altshuler1985} the $e-e$ scattering rate is 

\begin{equation}
\frac{1}{\tau_{ee}}=\frac{\pi(kT)^2}{4\hbar E_F} \cdot ln(\frac {\hbar q_sv_F}{kT})
\label{ee}
\end{equation}
, where $k$ is Boltzmann's constant, $E_F$ and $v_F$ are  Fermi energy and velocity and $q_s$  is screening wave vector. A comparison of the temperature dependence of the $\tau_q$ with the theory is shown in figure \ref{tau_q} by the solid black line. The comparison utilizes the screening wave vector $q_s$ as the only fitting parameter. The theory and experiments are in a good agreement. The screening waver vector is found to be  $q_s=3.38 \cdot 10^8$ (1/m), which is about 1.6 times higher than the Thomas-Fermi screening wave vector in 2D, given by $q_{TF}=2me^2/(\epsilon \hbar^2)=1.96 \cdot 10^8$ (1/m), where $\epsilon=12.9$ is the GaAs lattice dielectric function.

Current accuracy of the experiment at $T<$15 K does not allow to distinguish the exact $T^2$ dependence of the quantum scattering rate $1/\tau_q$ from the one given by eq.(\ref{ee}). The pure $T^2$ dependence is found theoretically for inelastic relaxation rate $1/\tau_{in}$ of non-equilibrium distribution function in Ref.\cite{dmitriev2005} (see eq. (37)  and eq.(42) there). The $T^2$ temperature dependence of the $1/\tau_{in}$ was observed in recent nonlinear experiments \cite{vitkalov2009}. The pure $T^2$ behavior of the inelastic relaxation is the result of a modification of electron screening in strong magnetic fields at a distance  $d \sim (D/\omega_c)^{1/2}$, where $D=(v_F)^2/(2\omega_c^2 \tau_{tr})$ is the diffusion coefficient in strong magnetic fields. At this scale there is a change in the dynamics of electron propagation from a ballistic motion at short distances $r<d$ to a "ballistic diffusion" at long distances $r>d$\cite{dmitriev2005}.  Shown in figure \ref{tau_q},  agreement between experiments and theory at zero magnetic field (see eq.(\ref{ee})) indicates that possible variations of electron lifetime $\tau_q$ due to the change of the electron dynamics in weak quantizing magnetic fields are small.  

On a qualitative level the obtained temperature dependence of the electron-electron scattering rate is consistent with results obtained in other experiments on different samples. Below  a quantitative comparison of the scattering rates obtained on the same sample by  different methods is presented.     

The comparison yields good agreement for the quantum scattering time $\tau_q$. Namely, the scattering time $\tau_q$=4.1 ps (see fig. 6a in Ref.\cite{vitkalov2009}) found from a numeric fit of a maximum of quantum oscillations is very close to the time $\tau_q$=4.15 ps, obtained from eq.(\ref{tau_q-T}) at temperature T=2.34 K.  

Shown in fig. \ref{tau_q} the temperature dependent part of the quantum scattering rate $1/\tau_q$:  $1/\tau_q(T)-1/\tau_q(T=0)=1.33 \cdot T^2$(GHz)  is stronger than a rate of inelastic relaxation of electron distribution: $1/\tau_{in}=0.56 \cdot T^2$ (GHz)\cite{vitkalov2009}. The rate was obtained in a nonlinear transport experiment on the same sample (see fig.6a in Ref.\cite{vitkalov2009}). The $T^2$ dependence indicates that the electron-electron scattering leads to the relaxation. A theoretical evaluation of the rate uses eq.(42) and eq.(37) in Ref.\cite{dmitriev2005}. The estimation yields $1/\tau_{in}^{th}=1.12 \cdot T^2$(GHz) at magnetic field $B$=0.15 (T) and $q_s=3.38 \cdot 10^8$ (1/m). This dependence is also weaker than the temperature dependence of the quantum scattering rate in fig. \ref{tau_q}.
 
The smaller rate of the electron inelastic relaxation $1/\tau_{in}$ is expected due to an ineffectiveness of the electron-electron scattering with a small momentum exchange for the relaxation of a specific electron distribution \cite{dmitriev2005}. The inelastic relaxation depends on the net amount of the electrons transferred in and out of each non-equilibrium part of the electron distribution by the electron-electron collisions. As shown in Ref.\cite{dmitriev2005}) the $e-e$ collisions with a small momentum exchange $q<1/d$  and with energy transfers that are multiples of cyclotron energy: $\Delta E \approx i \cdot \hbar \omega_c$ ($i$=0,1,2...  is integer) do not relax the non-equilibrium electrons, due to a periodicity of the non-equilibrium part of the electron distribution. The quantum lifetime $\tau_q$ of an electron is sensitive to any collisions, including the collision with the small momentum exchange. 

The theoretical estimation of the inelastic scattering rate is considerably higher the experimental value $1/\tau_{in}$. In Ref. \cite{vitkalov2009} an additional screening of electron-electron interaction  by X-electrons in buffer layers was suggested as a possible reason of the difference between the experiment and the theory. The present results indicate that if the additional screening exists it should enhance significantly the electron-electron scattering with the small momentum exchange in order to keep the overall (quantum) scattering rate to be intact.

\begin{figure}[t!]
\includegraphics[width=3in]{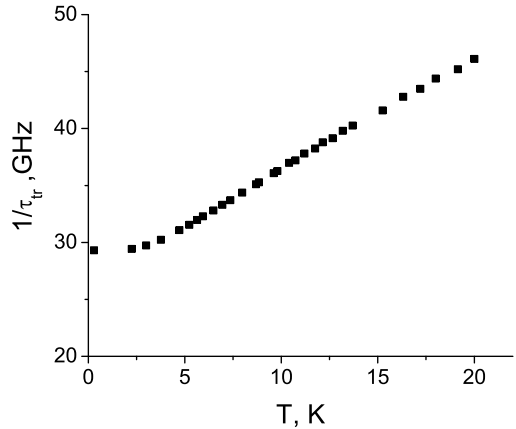}
\caption{Temperature dependence of transport scattering time $\tau_{tr}$. }
\label{tau_tr}
\end{figure}

Figure \ref{tau_tr} shows the temperature dependence of transport scattering rate $1/\tau_{tr}$. The dependence is obtained from the resistivity at zero magnetic field. At all temperatures the transport scattering rate $1/\tau_{tr}$ is much smaller than the quantum scattering rate $1/\tau_q$, indicating dominant contribution of the small angle scattering to the resistance even at high temperatures. The functional form of the temperature dependence of the transport scattering rate $1/\tau_{tr}$ is significantly different from the $T^2$ dependence of the quantum scattering rate $1/\tau_q$, shown in figure \ref{tau_q}. The main reason for the difference is that electron-electron scattering does not affect directly the electron transport, due to the conservation of the total momentum of colliding electrons. The electron collisions, nevertheless,  can transfer an electron from one quantum state to another states, decreasing the lifetime of the electron in a given quantum state.

----------------------------------------------------------------------
\section{Limitations of the method}

The comparison presented above  indicates very good agreement between experiment and theory\cite{vavilov2004}. Extracted from analysis, the quantum scattering rate $1/\tau_q$ is consistent with measurements  by other methods\cite{vitkalov2009}. The temperature variations of the electron-electron scattering rate  are in accord with the conventional theory \cite{chaplik,quinn,altshuler1985} and with  variations of the rate obtained from a nonlinear response on the same sample\cite{vitkalov2009}. 

Below we discuss reasons for complimentary agreement between experiment and the theory of the positive magnetoresistance as well as  possible limitations of the method. A difficulty of the practical implementation of the method to general systems is a contribution of other mechanisms to the magnetoresistance, which are beyond the theory\cite{vavilov2004}. The theory considers the disordered potential in, so-called, Self-Consistent Born Approximation (SCBA)\cite{uemura}. Within the SCBA the scattering events are completely uncorrelated.  

\begin{figure}[t!]
\includegraphics[width=3in]{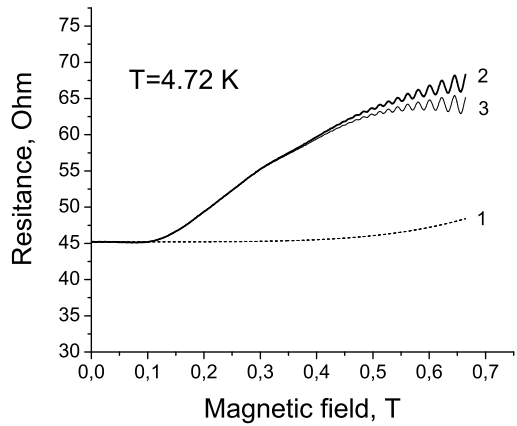}
\caption{Thin curve (1) shows contributions of correlated disorder and scattering to magnetoresistance. The curve (1) is plotted in accordance with eq.(\ref{polyakov}). Curve (2) presents experimental data for the magnetoresistance at $T=$4.72 K. Curve (3) in difference between curve (2) and curve (1), indicating negligibly small effect of the correlated disorder on the magnetoresistance below 0.3 Tesla. }
\label{class}
\end{figure}

Different theories, accounting for correlations in electron scattering (non-markovian or memory effects) and correlated disorder, indicate a wide-ranging variety of possible behavior of the magnetoresistance\cite{entin,shklov,polyakov}. A quantitative account of all possible effects may create difficulties for applications of the presented method, since some parameters significant for these theories may not be known {\it a priori}. A good practical indication of the small contribution of correlated disorder to the electron transport is the absence of the magnetoresistance at small magnetic fields, at which Landau levels are not formed yet ($\omega_c \tau_q \ll 1$). In accordance with classical (Drude) theory, which also ignores  correlations in electron scattering, the magnetoresistance must be absent in one valley conductors\cite{ziman}. This is the case for our samples (see figure \ref{class}). Indeed, the magnetoresistance is very small at $B<$0.07 Tesla.

Below we estimate a contribution of effects of the correlated scattering and correlated disorder to the magnetoresistance in our sample. As a first step we have to evaluate the correlation length of the disorder $\xi$. In high mobility samples the electrically charged donors are displaced from conducting layer by a distance $l$. Inside the conducting layer the displaced electric charges create a weak and smooth fluctuating electric potential with correlation length $\xi \approx l$. The weak potential induces the small angle scattering of electrons. An electron needs many scattering events to relax (randomize) it's original momentum. As a result the transport scattering time $\tau_{tr}$, describing the relaxation of the electron momentum, is much longer than the quantum scattering time $\tau_q$, which is an average time between two successive scattering events. Due to the scattering the direction of electron momentum performs a diffusive like motion with a typical step $\theta_0 \sim \hbar/(p_F \xi) \ll 1$, where $p_F$ is electron momentum. During the transport scattering time $\tau_{tr}$ an electron scatters about $N=\tau_{tr}/\tau_q$ times by the disorder potential and changes it's direction by $(\Delta \theta)^2 =\theta_0^2 \cdot N \sim 1$.  Thus  $\tau_{q}/\tau_{tr} \approx \theta_0^2=(\hbar/(p_F \xi))^2$ \cite{vavilov2004}. The ratio of these two times provides an estimation for the correlation length of the disordered potential $\xi$. In our samples  the quantum scattering time at T=2K is about 4 (ps)(see fig.3), whereas the sample conductivity at zero magnetic fields yields transport scattering time of $\tau_{tr}=$32 (ps). Thus the disorder correlation length is about $\xi \sim (\hbar/p_F) \cdot (\tau_{tr}/\tau_q)^{1/2} =12$ (nm). 

Figure \ref{class} presents effects of correlated disorder and correlated scattering on the magnetoresistance in our sample. In accordance with the theory a smooth disorder provides two distinct contributions to the resistivity \cite{polyakov}:

\begin{equation}
\frac{\Delta \rho_{xx}}{\rho_0}=-(\frac{\xi}{R_c})^2+ \frac{2\zeta(3/2)}{\pi}(\frac{\xi}{l_{tr}})^3(\frac{l_{tr}}{R_c})^{9/2}
\label{polyakov}
\end{equation}
, where $l_{tr}$ is mean free path of electrons moving in the smooth disorder and $R_c$ is cyclotron radius.  In eq.(\ref{polyakov}) the first term is due to a bending of electron trajectories within the correlation length $\xi$. The second term is associated with a classical memory effect due to the circular motion of electrons in a magnetic field \cite{mirlin1999}. In figure \ref{class} the line (1) shows the contributions of these effects to classical magnetoresistance, which is plotted in accordance with eq.(\ref{polyakov}) for correlation length $\xi=12$ (nm). A thick solid line (2) shows experimental results at T=4.72K. The thin solid line (3) is a difference between upper (2) and lower (1) curves, demonstrating the magnetoresistance without contribution of the classical memory effects.  Below B=0.3 T curves (2) and (3) are indistinguishable, indicating very small contribution of the correlated disorder to the resistivity. 

At B=0.15T the magnetic length $\lambda=66$ (nm) is considerably longer the correlation length of the disorder $\xi=12$ (nm) and SCBA works well for most part of scattering events. Thus the theory\cite{vavilov2004}, accounting for the interference contribution of returning paths  near this magnetic field, provides the leading contribution to the magnetoresistance.

\section{Conclusion}

The paper presents a simple transport method to access the electron lifetime $\tau_q$ of two-dimensional electrons in quantizing magnetic fields in broad temperature range. For two-dimensional electrons in GaAs quantum wells, the temperature variations of the quantum scattering rate $1/\tau_q$ are found to be proportional to the square of the temperature at $T<$15 Kelvin and are in very good agreement with the theory taking into account electron-electron interactions in 2D systems.

\begin{acknowledgements}

S. A. Vitkalov thanks Science Division of CCNY, CUNY Office for Research and NSF  (DMR 1104503) for the support of the experiments. A. A. Bykov thanks the Russian Foundation for Basic Research, project no. 11-02-00925.

\end{acknowledgements}


\begin{references}
\bibitem{prefactor} Equation (1) was derived in the absence of electron-electron interaction. While the exponential terms will likely contain a simple combination of quantum disorder scattering rate and electron-electron scattering rate, there is a possibility that  prefactor $\alpha$ in the last term of Eq. (1) depends  only on disorder scattering time. In the paper the parameter $\alpha$ is used as a single fitting parameter. The possible overestimation of the electron lifetime is  below 3\%.
\end{references}
\end{document}